# Three-Dimensional Structure of Hybrid Magnetic Skyrmions Determined by Neutron Scattering

WLNC Liyanage,[1] Nan Tang,[2] Lizabeth Quigley,[2] Julie A. Borchers,[3] Alexander J. Grutter,[3] Brian B. Maranville,[3] Sunil K. Sinha,[4] Nicolas Reyren,[5] Sergio A. Montoya,[4] Eric E. Fullerton,[4] Lisa DeBeer-Schmitt,[6] Dustin A. Gilbert[1,2*]

[1] Department of Physics and Astronomy, University of Tennessee
[2] Materials Science and Engineering Department, University of Tennessee
[3] NIST Center for Neutron Research, National Institute for Standards and Technology
[4] Center for Memory and Recording Research, University of California, San Diego
[5] Unité Mixte de Physique, CNRS, Thales, Université Paris-Saclay, Palaiseau 91767, France
[6] Neutron Scattering Division, Oak Ridge National Laboratory

*Corresponding Author: dagilbert@utk.edu

**Abstract**

Magnetic skyrmions are topologically protected chiral spin textures which present opportunities for next-generation magnetic data storage and logic information technologies. The topology of these structures originates in the geometric configuration of the magnetic spins – more generally described as the structure. While the skyrmion structure is most often depicted using a 2D projection of the three-dimensional structure, recent works have emphasized the role of all three dimensions in determining the topology and their response to external stimuli. In this work, grazing-incidence small-angle neutron scattering and polarized neutron reflectometry are used to determine the three-dimensional structure of hybrid skyrmions. The structure of the hybrid skyrmions, which includes a combination of Néel-like and Bloch-like components along their length, is expected to significantly contribute to their notable stability, which includes ambient conditions. To interpret the neutron scattering data, micromagnetic simulations of the hybrid skyrmions were performed, and the corresponding diffraction patterns were determined using a Born approximation transformation. The converged magnetic profile reveals the magnetic structure along with the skyrmion depth profile, including the thickness of the Bloch and Néel segments and the diameter of the core.

**Keywords**: skyrmions; neutron scattering; micromagnetic modeling

**Introduction**

Chiral spin textures in magnetic thin films and their resulting topology are currently at the forefront of theoretical and experimental condensed matter research due to their potential for topologically protected data storage and logic devices and as well as their exciting fundamental physics.[1-5] These structures can generally be described as a local whirl of the spin configuration, forming continuous closed features.[4] The continuous wrapping of the spin structure bestows a non-trivial topology, which stabilizes them from certain types of distortion, and as a result skyrmion data technologies are inherently resistant to data corruption and degradation. Magnetic skyrmions have been the most prevalent of these topological structures and consists of a continuous, coplanar wrapping of the magnetic moments with the moments at the core and perimeter of the wrapped structure oriented in the out-of-plane direction with opposite orientations. In conventional skyrmions this structure is continuous throughout the thickness of the material, forming a skyrmion tube. Typically, magnetic skyrmion tubes order into hexagonally packed 2D lattices with the wrapping in the plane of the lattice. These spin textures were first observed in B20-structured MnSi single crystals using small-angle neutron scattering (SANS)[6], and later in FeGe, MnFeGe, Fe$_{1-x}$Co$_x$Si, and Cu$_2$OSeO$_3$, among others.[7-9] In these systems, the non-centrosymmetric crystal structure results in a non-zero Dzyaloshinskii-Moriya interaction (DMI), which causes the magnetic moments to curl into a variety of chiral structures with fixed handedness defined by the sign of the DMI . Recent skyrmion research has increasingly focused on thin-film systems, including thin-film B20 materials and skyrmions stabilized by interfacial DMI.[5,7,10,11] However, few skyrmion systems are able to stabilize skyrmions at ambient conditions (zero field and room temperature), which is necessary for the deployment of skyrmions in consumer technologies.[1,12-14]

For conventional skyrmions, the in-plane wrapping can be Néel-type, with the magnetic moments curling along the radial direction between the core and the perimeter, or Bloch-type, with the moment curling in the azimuthal direction, as described previously.[10] Recently, skyrmions which are stable across a wide range of temperatures and magnetic fields (10 K – 325 K; -30 mT – +200 mT) have been reported in Gd/Fe multilayers.[1] These skyrmions are proposed to have a hybrid structure consisting of both Bloch and Néel-type components.[15-17] The hybrid skyrmion can be considered in a thin film as a Bloch-type skyrmion with flux closure domains at the ends of the tube, i.e. where the skyrmion tube meets the top and bottom surfaces of the film; the flux closure domains are in-plane structures which act to vector the dipolar fields into the film, reducing the magnetostatic energy.[18,19] These flux-closure domains follow the wrapped contour of the skyrmion tube, themselves forming a closed wrapping, making them equivalent to a Néel-type skyrmion. Since the in-plane component of the dipolar fields are in opposite directions on the top and bottom surface, the flux-closure domains will have opposite chiralities. Thus, a hybrid skyrmion can be generally described as a three-dimensional stack of three skyrmions, evolving from Néel-type at one end of the skyrmion tube to Bloch-type, and back to Néel-type at the opposite end of the skyrmion tube, with the two ends having opposite chiralities.[1,15,16,20] Also notable, while conventional skyrmions are stabilized by the Dzyaloshinskii-Moriya interaction, skyrmions in the Gd/Fe system are stabilized by dipolar interactions; since the flux-closure domains, e.g. Néel-type caps, also shape the dipolar fields, the stability, which has been demonstrated as exceptional, is directly related to the structure.

The three-dimensional structure of magnetic skyrmions is critical to their topology, stability, and their dynamic properties when perturbed with external forces. Verifying and quantifying their three-dimensional structure is important to understand how these complex magnetic spin textures exist as localized magnetic objects in magnetic materials. Previous works have used electron holography to determine the three-dimensional structure of conventional Bloch-type magnetic skyrmions in nanowires, at dimensions limited by the transmission of electrons through the material.[21] For hybrid skyrmions in particular, the requirement for a significant dipolar interaction means these features preferentially manifest in thin films and would likely be unstable in nanowires. To better understand the stability of hybrid skyrmions, it is critical to determine their three-dimensional structure.

Addressing the need for *in situ* probes of hybrid skyrmion structure requires a penetrative, non-destructive tool such as neutron scattering,[22,23] which has been used extensively in skyrmion research due to its sensitivity to nanoscale magnetic features. However, the vast majority of these investigations have been performed on bulk materials due to limitations in signal strength and lack of skyrmion thin-films with long-range coherent order [24]. Hybrid skyrmions observed in Fe/Gd multilayer thin-films have been previously measured with transmission SANS, which captures the in-plane hexagonal ordering of the skyrmion lattice but not its depth profile.[1] Alternatively, polarized neutron reflectometry (PNR), which is typically performed in a specular reflection geometry, provides access to the structural and magnetic depth profile, but only the depth-resolved average; in-plane structure is generally not probed with specular PNR.[25] By combining these techniques in an appropriate scattering geometry, grazing-incidence small-angle neutron scattering (GISANS) and off-specular PNR allow simultaneous access to the in-plane and out-of-plane structures.[26-28]

In this paper, we perform PNR, SANS, and GISANS measurements to determine the three-dimensional structure of hybrid skyrmions in Fe/Gd multilayers.[1] First, SANS and PNR measurements are performed on a single thin-film to determine the structure of the hybrid skyrmion. The results from these scattering experiments are then used to inform and constrain a micromagnetic simulation. Using a Born approximation algorithm, the GISANS pattern from the simulation is calculated and compared to experimental GISANS data. Through iterative feedback between the experimental and simulated results, the micromagnetic simulation is adjusted to reproduce the experimental results. Our combined experiments and micromagnetic model reveal the complex three-dimensional structure of the hybrid skyrmion in a thin film magnet.

**Experimental Procedure**

Thin-film multilayers with a nominal structure of $[(\text{Fe } (3.6 \text{ Å})/\text{Gd } (4.0 \text{ Å})]_{120}$ were grown on a 1 cm × 1 cm Si (001) substrate with native oxide layer by DC magnetron sputtering, as described previously[1,15,16]. A Ta (5 nm) seed and capping layer was used for adhesion and protection against oxidation, respectively. X-ray diffraction shows no apparent peaks, indicating the film is amorphous or nanocrystalline, consistent with previous results. Magnetic hysteresis loops were measured and shown in the Supplemental Material.[29] The layered structure of the film induces a modest perpendicular magnetic anisotropy. The skyrmion state is prepared following a previously published field sequence[1]: first, the sample was saturated by a 500 mT magnetic field applied at 45° relative to the out-of-plane direction and then returned to remanence (0 mT). After returning to remanence the magnetic configuration is comprised of oriented stripe domains. Next, the magnetic field is increased along the out-of-plane direction; at 190 mT, the magnetic domains pinch to become isolated skyrmions in an ordered, oriented hexagonal lattice.[1] The magnetic field was then removed, and the system returned to remanence.

Polarized neutron reflectometry measurements were performed using 5 Å wavelength neutrons, at room temperature, on the MAGIK reflectometer at the NIST Center for Neutron Research (NCNR). Measurements were performed in the saturated state with a 500 mT in-plane field, and in the skyrmion state with a small (≈1 mT) out-of-plane guide field. By the scattering selection rules, the reflectivity is sensitive only to the projection of the magnetization in the sample plane that is perpendicular to the scattering wavevector, $Q$.[30] The saturated state provides magnetic and nuclear information in the non-spin flip channel ($R^{++}$, $R^{--}$, where ++ (--) indicates the incident and scattered neutron spin orientation as up-up (down-down), respectively). Specifically, the splitting between the $R^{++}$ and $R^{--}$ reflectivities in this scattering configuration originates in the projection of the net in-plane magnetization that is parallel to the applied field. In contrast, the out-of-plane guide field for the skyrmion state causes the magnetic information to manifest only in the spin-flip channel, $R^{+-}$ and $R^{-+}$ (where +- (-+) indicated the incident and reflected neutron spin orientation as up-down (down-up))[31], and $R^{++}$ and $R^{--}$ have contributions only from the structural scattering. Data fitting was performed using the Refl1d software package [32]; the models for the two field conditions were fitted in parallel, with the nuclear parameters coupled. The specular PNR technique is sensitive to the depth-resolved, average, in-plane magnetization, and is insensitive to in-plane magnetic regularity or out-of-plane magnetization.[25,27]

SANS and GISANS measurements were performed on the very small-angle neutron scattering (vSANS) instrument at the NIST Center for Neutron Research using 14 Å neutrons at room temperature. The SANS measurements were performed with the neutron beam parallel to the film's out-of-plane direction and near-zero magnetic field, resulting in the diffraction pattern for the stripe (Fig. 1a) and skyrmion states (Fig. 1b). Grazing-incidence SANS measurements were performed by illuminating the sample at grazing incidence, similar to PNR but now on a SANS instrument. The incident angle was varied, between 0.3° to 0.9° in steps of 0.05°. An illustrative diagram of the GISANS measurement setup is shown in Fig. 1c. Unlike PNR, GISANS captures the specular reflection, as well as any off-specular reflections from in-plane ordering. A small magnetic field is applied in the film's out-of-plane direction during the measurement. The sample was measured in the stripe, skyrmion and saturated states. Data are presented in terms of the momentum transfer vector, with $Q_z$ defined as the out-of-plane direction.

Micromagnetic simulations were performed using the Object Oriented Micro Magnetic Framework (OOMMF) simulation platform. [33] This platform uses a macrospin approximation with cells 5 nm in the x and y direction and 1 nm in the z-direction; this length is smaller than the calculated exchange length. The total simulation area was 880 nm × 760 nm nm in the *x* and *y* directions, respectively, with a thickness in *z*-direction based on the PNR-determined depth profile. Magnetostatic, magnetocrystalline, and exchange energies are considered in the model with periodic boundary conditions, and no net DMI is expected in the experimental system. The anisotropy constant is determined from magnetometry using the in-plane saturation field and ferromagnetic resonance studies, and the saturation magnetization is determined using the experimental PNR results, measured at saturation (500 mT in-plane field). The simulations were allowed to relax to a steady state with zero applied magnetic field. Using the Born approximation and an in-house python code, the diffraction pattern is calculated from the OOMMF simulation for any scattering plane and neutron spin orientation.[25]

**Results**

SANS measurements were performed on single films of Fe/Gd multilayers in the stripe and skyrmion states. The SANS pattern measured in the stripe phase, Fig. 1a, shows two peaks located at $|Q| = \pm 0.00339$ Å$^{-1}$, corresponding to a real-space periodicity of 185 nm. The presence of only two peaks indicates that the stripes have a long-range orientation, as opposed to labyrinth domains which would show a uniform intensity ring-like SANS pattern [34]. The position of the peaks confirm that the long axes of the stripes are parallel to the direction of the in-plane field component during the preparation of the initial state. Subsequently preparing the sample in the skyrmion state, the SANS pattern has a six-peak hexagonal structure shown in Fig. 1b, representing the Fourier transform of the hexagonally-ordered skyrmion lattice within the film. As in the stripe phase, the hexagonal pattern in the skyrmion state demonstrates that the skyrmions are arranged in a well-ordered lattice which is highly oriented across the film plane. The peaks in the hexagonal pattern at zero field and room temperature are evident at $|Q| = 0.00298$ Å$^{-1}$, corresponding to a skyrmion center-to-center spacing of 211 nm. These data provide insight into the in-plane spatial ordering of the film, but do not provide details pertaining to the depth-dependent structure of the hybrid skyrmions through the film.

Complementary to the SANS measurements, PNR measurements provide details about the depth-resolved average structure. Neutron reflectometry results and the converged depth profile from the saturated and skyrmion state measurements are shown in Figs. 2(a, b) and 2(c, d), respectively. Since signal-to-noise limitations obscure the superlattice diffraction peaks from the [(Fe (3.6 Å)/Gd (4.0 Å)]$_{120}$ bilayers in these measurements, the multilayer structure is treated as a single composite Fe$_x$Gd$_{1-x}$ layer with a nominal x=0.47. The converged depth profile confirms the structure of the film to be Si / Ta(6.4 nm) / FeGd(81.7 nm) / Ta(5.3 nm) / TaO$_x$(2.2 nm) and provides an experimental nuclear scattering length density (nuclear scattering length density, SLD, $\rho_N$) $\rho_N$(Ta) = 4.13×10$^{-6}$ Å$^{-2}$, $\rho_N$(GdFe) = 3.76×10$^{-6}$ Å$^{-2}$. These values are slightly different from the expected values for the bulk constituents ($\rho_N$(Ta) = 3.83×10$^{-6}$ Å$^{-2}$, and $\rho_N$(GdFe) = 4.23×10$^{-6}$ Å$^{-2}$) likely indicating some oxidation of the Ta and that the Gd layer thickness may be less than the nominal value. The Gd/Fe ratio can be uniquely identified by the imaginary SLD, $\rho_{Imag.}$, and corresponds to neutrons being absorbed by the film, which occurs exclusively in the Gd; no apparent critical edge is observed in the reflectometry due to the absorbing nature of the Gd.

The PNR data measured in the saturated state are shown in Fig. 2a, with the converged profile in Fig. 2b; the reflectivity from the model is shown as a solid line in Fig. 2a. The profile provides a magnetic SLD ($\rho_M$) which confirms that only the FeGd layer is magnetic, and it is uniformly magnetic throughout the entire thickness of the thin-film, with $\rho_M$(FeGd) = 0.99×10$^{-6}$ Å$^{-2}$. The magnetic SLD can be transformed to a magnetization of 350 kA m$^{-1}$ (1 kA m$^{-1}$ = 1 emu/cm$^3$). Since Gd and Fe have an antiferromagnetic exchange interaction, this is the net magnetization from the antiparallel layers, as the applied magnetic field is insufficient to break the antiparallel coupling.[35] The saturation magnetization and structural parameters are used later to inform and constrain the micromagnetic simulations.

Reflectometry data taken in the skyrmion state are shown in Fig. 2c, with the fitted magnetic depth profile shown in Fig. 2d. The converged model shows a small in-plane magnetization through the middle of the film, with a slightly larger (≈20%) magnetization at the top and bottom surfaces. While an initial consideration of the PNR profile of a skyrmion would suggest that there should be no magnetic scattering, since the average in-plane magnetization for a skyrmion is zero, previous result have shown that this assumption is not correct. Specifically, in PNR the neutron wave packet is reflected off the surface of the sample at a shallow angle, experiencing the average scattering potential from the sample within a volume defined by the longitudinal (i.e. at the intersection of the scattering and sample planes) and transverse (i.e. in the sample plane but perpendicular to the scattering plane) neutron coherence length. The longitudinal neutron coherence length in the reflectometers at NIST has been measured to be ≈100 μm, however the transverse coherence length is <1 μm.[36] If the transverse coherence length is not an integer multiple of the skyrmion periodicity, the neutron will interact with a non-integer number of skyrmions, i.e. catching the edge of a skyrmion. This extra part of the skyrmion within the neutron coherence length contributes a net in-plane magnetization.[25] Since an out-of-plane guide field is used, the orientation of these fractional skyrmions is irrelevant: they are all orthogonal to the neutron spin and contribute similarly to the spin-flip signal. This is important because it means that opposite edges of the skyrmion and their winding (Bloch or Néel) all contribute to the average. As a result, the PNR spin-flip signal is non-zero and possesses a sensitivity to the in-plane magnetization. Similar discussions were presented previously in Ref. [25]. Based on this understanding, peaks in the magnetization near the top and bottom of the GdFe layer in the skyrmion state, shown in Fig. 2d, suggest that the Néel caps have a larger in-plane footprint than the Bloch-type wrapping around the equatorial band. The thickness of the top and bottom magnetization peaks provides the first insight into the thickness of the Néel caps, which we can identify here as approximately 17.9 nm (1σ confidence interval of 11.9 nm – 18.4 nm).

## GISANS

Grazing-incidence SANS, similar to off-specular reflectometry, provides simultaneous insight into the in-plane and out-of-plane structure. As noted above, GISANS measurements were taken at a series of angles ($\theta$) between 0.3° to 0.9° in steps of 0.05°; where $\theta$ corresponds to the angle between the sample plane and the neutron beam (Fig. 1c). At each angle, the 2D scattering pattern was captured, as shown in Fig. 3a for $\theta$ = 0.5° in the stripe phase. In this state, three GISANS peaks can be observed, one at finite $Q_z$ and $Q_y$ = 0, representing the specular reflection, and two at $Q_y$ = ±0.00338 Å$^{-1}$, representing the off-specular diffraction. The peak location along the $Q_z$ coordinate corresponds to the traditional 2$\theta$ position in standard reflectometry or diffraction geometry and changes with the sample angle. The peak location along $Q_y$ is determined by the in-plane periodicity, which for the stripe phase is 186 nm. The stripe state corresponding to zero applied field has two peaks analogous to the transmission SANS pattern shown in Fig. 1a.

Compared to the stripe domain configuration, the specular scattering in the skyrmion state is nearly identical while the off-specular scattering intensity is much weaker and has a narrower distribution along the $Q_z$ direction, as shown in Fig. 3b. The off-specular peaks can be observed at $Q_y$ = 0.0025 Å$^{-1}$ when measured in a magnetic field $\mu_0 H$ = 190 mT, corresponding to a real space periodicity of 251 nm. Our previous work[1] showed that in these hybrid skyrmions, the periodicity increased at larger fields following a soliton lattice theory.[37] To further clarify that these peaks are solely from the magnetic structure, GISANS measurements were performed in the saturated state using an out-of-plane magnetic field. The external field pulls the magnetic moments in the z direction (out-of-plane direction). According to the neutron selection rules, the contribution to the magnetic scattering at the specular position should be zero if the sample is saturated. The magnetic peaks at finite $Q_y$ should be evident if the skyrmion lattice persists. Instead, these peaks are gone, which indicates off-specular peaks were magnetic in origin, as shown in Fig. 3c. We may therefore reliably conclude that the off-specular scattering is exclusively magnetic in nature.

Preliminary analysis of the data can be performed by reducing the specular data from the GISANS measurements to a 1D intensity versus $Q_z$ plot and comparing this with the PNR results. As noted above, the specular reflection in the saturated state, measured with an out-of-plane guide field, captures the nuclear structure, which was also measured in the PNR measurements ($R^{++}$ and $R^{--}$ in Fig. 2c). The converged nuclear depth profile from Fig. 2d is used, with $\rho_M$ set to zero, to generate the corresponding nuclear-only reflectometry with GISANS instrumental conditions; this reflectometry pattern is compared to the 1D GISANS results in Fig. 3d and shown to agree well, with a chi-squared ($\chi^2$) of 1.21. Alternative models were tested and produced poorer fits, with a 'flat' magnetic structure resulting in a $\chi^2$ of 2.87 and a non-magnetic structure 2.07. To analyze the off-specular peaks, a physics-based model of the magnetization is generated using the OOMMF simulation platform; the GISANS pattern from the OOMMF result is calculated and compared to the experimental results.[25]

The intensity of the off-specular peaks is integrated along $Q_y$ and used to generate a 1D plot of the of the off-specular intensity as a function of $Q_z$, shown in Figure 3e. Since this off-specular scattering appears at finite values of $Q_z$ and $Q_y$, these data encode both the in-plane and out-of-plane structure, respectively, of the hybrid skyrmions. The next section discusses the micromagnetic model used to generate the simulated data in Figure 3e.

**Micromagnetic modeling**

Hybrid skyrmions were simulated using physical parameters derived from the experimental results. The saturation magnetization was determined from magnetization measurements to be 351 kA m$^{-1}$; the uniaxial anisotropy constant ($K_U$) was determined to be $1.8\times10^5$ J m$^{-3}$ using the in-plane (hard-axis) saturation field.[38] The lateral exchange stiffness ($A$) was adjusted to achieve a stable skyrmion configuration with a periodicity that matched the experimental SANS results, converging to a value of $0.7\times10^{-11}$ J m$^{-1}$. This parameter is expected to be much less than iron ($2.1\times10^{-11}$ J m$^{-1}$)[39] due to the nearly 2D nature of the film layers and was the primary control variable in the modeling. For reference, this value is slightly less than nickel ($0.9\times10^{-11}$ J m$^{-1}$) and is consistent with ferromagnetic resonance measurement in Ref. [15]. The vertical exchange stiffness was fixed at 10% of the lateral, reflecting the difference between Fe-Fe exchange coupling and Fe-Gd exchange coupling.[35] The initial state of the system was prepared with hybrid skyrmions, which has Néel caps at the ends of the skyrmion tube and a Bloch-type winding at/near the equatorial band. Then the micromagnetic simulation was allowed to relax in zero field. For the case of too large or small $A$, $K_U$ or $M_S$, the state collapsed to labyrinth domains or a saturated state. Including a bulk DMI term results in ejection of the Néel-type surface states, resulting in traditional Bloch-type tubular skyrmions; including DMI at the surfaces or interfaces may result in a Néel type skyrmion, as is common in many multilayer systems. The model was also tested by starting with only Bloch or Néel-type states, or Néel-type (flux closure caps) with the same chirality, all of which relaxed to the hybrid configuration or collapsed to a saturated or labyrinth state, confirming the robustness of the hybrid skyrmion configuration. The simulation space (880 nm × 760 nm × $t$, where $t$ is the layer thicknesses determined from PNR) captures an array of 16 skyrmions.

A plane view image and a series of real-space slices of a single hybrid skyrmion from the converged model are shown in Fig. 4(a-f). This simulation, which includes all of the underlying physics of the magnetic interactions, confirms the proposed structure, including the Néel-type surface states with opposite chirality at the top and bottom portion of the thin-film, and a Bloch-type winding at/near the equatorial band. Comparing Figs. 4b and 4f with 4c-e, we see that the width of the skyrmion boundary is larger in the Néel caps (b and f) and narrower in the Bloch region (c-e). The in-plane width of the skyrmion boundary is extracted by fitting a Gaussian function to the in-plane magnetization and determining the width, and is plotted in Fig. 4g. The width of the boundary is shown to increase from 16 nm at the equatorial band to 22 nm in the Néel caps. Since the in-plane magnetization detected by PNR originates within this in-plane boundary region, a larger skyrmion boundary width yields a larger effective in-plane magnetization in PNR. To understand this scaling, Fig. 4(h) plots the total in-plane magnetization at each depth ($1 - \int |M_z| \, dxdy$) alongside the magnetic SLD from the PNR results. We note that while the spin-flip PNR is sensitive to the in-plane magnetization averaged over the neutron coherence length rather than the *total* in-plane magnetization as defined above, the two signals are expected to be approximately linearly related to each other in this case. As discussed above, the magnetic SLD detected by PNR in this system is governed by the coherence volume of the neutron wave packet, which overlaps with a non-integer number of skyrmion boundaries such that a residual in-plane magnetization is detected.

Comparing the OOMMF model to the Refl1D model, the structures are very comparable. The nuclear structures are nearly identical by design and are thus not shown. The magnetic profiles of both models similarly show a larger magnetic feature near the surface. The thickness of the feature is 13 nm in the OOMMF model and 17.9 nm (1σ confidence interval of 11.9 nm – 18.4 nm) in the Refl1D model. Notably, the OOMMF model is within the confidence interval of the experimental data. The key difference in the two magnetic profiles is the sharpness of this feature, more specifically the interface between the Bloch and Néel regions. The OOMMF model shows a wide boundary, spanning nearly 30 nm, while the Refl1D model has a much sharper interface of 4.3 nm ± 2.3 nm. The Refl1D model has poor sensitivity to this feature since the large width of this magnetic interface will generate exceedingly weak reflectometry fringes for larger interface widths.

Another interesting insight provided by the OOMMF results is the continuous evolution of the Néel/Bloch/Néel structure with depth. Specifically, Figures 4(b) and (f), which correspond to the ends of the skyrmion tube and should be Néel-type, are not entirely oriented along the radial direction in the boundary region. Instead they are canted by an average of 13° towards the azimuthal direction. Figures 4(c) and (e), which are between the Bloch and Néel regions, show that the spin moments are oriented between the radial and azimuthal directions. The equatorial band of the skyrmion tube, Figure 4(d), is oriented along the azimuthal direction, and is therefore a proper Bloch structure. The angle of the magnetic moment in the boundary region versus depth is extracted and plotted in Figure 4(i). This plot shows that the angle of the spins, relative to the azimuthal (in-plane) direction, changes continuously throughout the film. While PNR is sensitive to the width of the boundary region with in-plane spins, the angular orientation of the moments within the plane cannot be distinguished in our scattering geometry. Interestingly, while the skyrmion boundary width has three distinct regions, which we above identify as the surface Néel caps and the Bloch region, the spin angle, which is the actual identifier of the Néel and Bloch regions, e.g. the in-plane angle of the magnetization in the boundary, is more continuous.

Using a Born approximation-based Python code, the PNR, SANS, and GISANS diffraction patterns from this structure were generated from the converged micromagnetic profile. The code included the real and imaginary nuclear structure as determined by PNR, and the magnetic structure determined by the micromagnetic model, constrained by the PNR-determined nuclear structure and magnetization. The roughness between each layer in the modeled system (e.g., Si/Ta, Ta/GdFe, GdFe/Ta, Ta/TaO, TaO/air) was coarsely treated as a 1 nm thick boundary layer with an SLD defined as the average of the adjacent layers; this limitation is set by the cell size in the micromagnetic model. Also included in the code was the spin of the neutron, defined to be along the film normal direction, matching the experimental measurements, and in the scattering plane. For comparison to the data in Fig. 1b, a preliminary simulation in the SANS (transmission) geometry was calculated and is shown in Fig. 5a. The structure confirms the hexagonal pattern which also appears in the simulated structure. Additional peaks and stretching along the $Q_y$ direction are the result of the finite volume of the simulation and fractional skyrmions on the boundary of the simulation space. The calculated diffraction peaks appear at $|Q| = 0.0033$ Å$^{-1}$, consistent with the experimental results, which showed $|Q| = 0.00298$ Å$^{-1}$, confirming the simulation captures the mesoscopic structure.

Next, the GISANS pattern, for this structure was calculated along the $Q_y$ and $Q_z$ directions, as shown in Fig. 5b. The simulated result is calculated at a single reflection angle to emulate the experimental result, and indeed generates a similar structure to Fig. 3b. The reflectometry data – which is operationally identical to a GISANS measurement captured at $Q_y$=0 – and GISANS data were measured at a series of sample angles then integrated to establish curves versus $Q_z$; an integrated series of GISANS measurements is shown in Fig. 5c. Both at the specular reflectometry ($Q_y$=0) and off-specular position ($Q_y$=0.0033 Å$^{-1}$) the Kiessig fringes are clearly visible, indicating a sensitivity to the depth profile. From these data, all four scattering cross-sections ($R^{++}$, $R^{--}$, $R^{-+}$, $R^{+-}$) are extracted at $Q_y = 0$ and $Q_Z<0$ (corresponding to a front-side reflection), generating a model of the specular reflectometry pattern based on the OOMMF model which is comparable to the experimental data and Refl1D fit in Fig. 2c. The experimental, Refl1D and OOMMF models are compared in Fig. 5d. The non-spin flip signal reproduces all of the major features and structure in both the Refl1D and OOMMF models, which is expected because it is based on only the nuclear structure, which is programmed to be nearly identical. The spin-flip data are comparable, with a notable matching of the higher frequency oscillations. Notable differences in the spin-flip signal can be seen by a more rapid decay in the Refl1D model and a damping of the high-frequency oscillations at $Q_z$=0.075 Å$^{-1}$. The former feature is likely a result of the background subtraction, and a limited ability to model roughness in the OOMMF model. The latter feature is likely a consequence of the small surface domains having a much sharper interface in the Refl1D model (converged to be 4.3 nm ± 2.3 nm) compared to the OOMMF model, which changes continuously over nearly 30 nm. The Refl1D has poor sensitivity to this feature since the large width of this magnetic interface will generate exceedingly weak reflectometry fringes.

The off-specular features from the GISANS data are also calculated and compared to the experimental results in Fig. 3e; these calculated data are the sum of the $R^{++}$, $R^{--}$, $R^{+-}$, and $R^{-+}$ to reproduce the experimental results. Compared to the PNR range, the GISANS range is limited because the detector must be located at the maximal distance in the SANS tube and measured with a long neutron wavelength to separate the off-specular signal from the specular reflection. However, in this configuration only a very small range of sample angles can be measured before the beam is off the edge of the detector. Nevertheless, the calculated results accurately reproduce the experimental data in Fig. 3e. Furthermore, these data capture the critical fall-off region in-which the neutrons penetrate the sample and interact strongly, and also present maximal neutron fluence and maximal statistical importance.

Together the OOMMF and Refl1D models are complementary. Both propose to determine the structure of the skyrmion, with the Refl1D approach utilizing scattering with no consideration for the underlying magnetic energies, while OOMMF has no ability to experimentally confirm its predicted structure but includes these magnetic interactions. This work has shown that the Refl1D model accurately captures key features of the system, such as the thickness of the surface state, but misses others, such as the interface width of this feature. OOMMF also provides insights which cannot be captured by scattering, such as the rotation angle of the in-plane moments, or their correlation with the surface states. Together, these results provide a structure that is grounded both in fundamental physics and confirmed through experimental investigation.

## Discussion

Hybrid skyrmions have been demonstrated as one of the few chiral magnetic systems which are stable at room temperature and zero magnetic field, making them a strong candidate for skyrmion-based technologies. Their exceptional stability can be attributed to their hybrid structure, specifically the flux-closure domains on the ends of the skyrmion tubes, that decrease the magnetostatic energies in the system. The structure is also key to the topology of the system and its dynamic response for high frequency applications. The present work experimentally determined the three-dimensional structure of the hybrid skyrmion using a combination of neutron scattering techniques and physics-based modeling. Using PNR and SANS, the in-plane and depth-averaged magnetic structure is determined. Details from these measurements are used to inform and constrain a physics-based micromagnetic model, which was iteratively improved to reproduce the experimental results. The micromagnetic model was subsequently verified by comparing with grazing incidence SANS results, which simultaneously captures the in-plane and out-of-plane structure. The real-space spin structure, provided by the micromagnetic model, first verifies the three-dimensional construction of the hybrid skyrmion and provides new insights about their structure. Key among these features is that the thickness of the Néel caps (13 nm – 18 nm), the continuous winding of the in-plane angle, and the changing width of the in-plane skyrmion boundary. This last observation is due to the role of the Néel caps as a flux-closure structure and may indicate an approach to increase skyrmion stability by designing heterostructures with artificial soft layers. Understanding this structure may allow other skyrmion materials to be augmented in such a way as to increase their stability, or alternatively, design the topology in artificial magnetic systems. The approach used here ties together the observed neutron scattering results to magnetic configurations which are constrained by their underlying physics, providing an improved investigative paradigm which can be leveraged to achieve an enhanced understanding of other magnetic systems.

## Acknowledgements

Neutron experiments and data analysis, including modeling and programming, were supported by the U.S. Department of Energy, Office of Science, Office of Basic Research Early Career program under Award Number DE-SC0021344. Work at UC San Diego was supported by the National Science Foundation, Division of Materials Research Award #: 2105400. Measurements at NIST were performed on the vSANS instrument which is supported by the Center for High Resolution Neutron Scattering, a partnership between the National Institute of Standards and Technology and the National Science Foundation under Agreement No. DMR-2010792. We appreciate the assistance of the NIST researchers who supported this work including Dr. Kathryn Krycka, Dr. Cedric Gagnon, and Dr. John Barker. A portion of this research used resources at the High Flux Isotope Reactor, a DOE Office of Science User Facility operated by the Oak Ridge National Laboratory. Sergio A. Montoya acknowledges support from the Department of Defense.

## Author Contribution

Project and experiment design was performed by D.A.G., J.A.B., L.D.-S., S.A.M. and E.E.F. Samples were prepared by S.A.M. and E.E.F. SANS and reflectometry experiments were performed by W.L.N.C., N.T., L.Q, L.D.-S., J.A.B., A.J.G., and D.A.G. OOMMF simulation were performed by W.L.N.C., L.Q. and D.A.G. The Python code was written by B.B.M., W.L.N.C., and D.A.G. Analysis of the results was performed by W.L.N.C., D.A.G., J.A.B., L.D.-S. S.S. and E.E.F. The first draft was written by W.L.N.C., and D.A.G. All authors made contributions to and approved the final text.

## References

[1] R. D. Desautels, L. DeBeer-Schmitt, S. A. Montoya, J. A. Borchers, S.-G. Je, N. Tang, M.-Y. Im, M. R. Fitzsimmons, E. E. Fullerton, and D. A. Gilbert, Realization of ordered magnetic skyrmions in thin films at ambient conditions, Phys. Rev. Matter. **3** (2019).

[2] D. A. Garanin, D. Capic, S. Zhang, X. Zhang, and E. M. Chudnovsky, Writing skyrmions with a magnetic dipole, J. Appl. Phys. **124** (2018).

[3] J. Chen, W. P. Cai, M. H. Qin, S. Dong, X. B. Lu, X. S. Gao, and J. M. Liu, Helical and skyrmion lattice phases in three-dimensional chiral magnets: Effect of anisotropic interactions, Sci. Rep. **7**, 7392 (2017).

[4] B. Göbel, I. Mertig, and O. A. Tretiakov, Beyond skyrmions: Review and perspectives of alternative magnetic quasiparticles, Phys. Rep. **895**, 1 (2021).

[5] W. Jiang, G. Chen, K. Liu, J. Zang, S. G. E. te Velthuis, and A. Hoffmann, Skyrmions in magnetic multilayers, Phys. Rep. **704**, 1 (2017).

[6] S. Mühlbauer, B. Binz, F. Jonietz, C. Pfleiderer, A. Rosch, A. Neubauer, R. Georgii, and P. Böni, Skyrmion Lattice in a Chiral Magnet, Science **323**, 915 (2009).

[7] Y. Zhou, Magnetic skyrmions: intriguing physics and new spintronic device concepts, Natl. Sci. **6**, 210 (2019).

[8] X. Z. Yu, N. Kanazawa, Y. Onose, K. Kimoto, W. Z. Zhang, S. Ishiwata, Y. Matsui, and Y. Tokura, Near room-temperature formation of a skyrmion crystal in thin-films of the helimagnet FeGe, Nat. Mater. **10**, 106 (2011).

[9] S. Heinze, K. von Bergmann, M. Menzel, J. Brede, A. Kubetzka, R. Wiesendanger, G. Bihlmayer, and S. Blügel, Spontaneous atomic-scale magnetic skyrmion lattice in two dimensions, Nat. Phys. **7**, 713 (2011).

[10] A. Fert, N. Reyren, and V. Cros, Magnetic skyrmions: advances in physics and potential applications, Nat. Rev. Mater. **2** (2017).

[11] A. Fert, V. Cros, and J. Sampaio, Skyrmions on the track, Nat. Nanotechnol. **8**, 152 (2013).

[12] K. Karube, J. S. White, D. Morikawa, M. Bartkowiak, A. Kikkawa, Y. Tokunaga, T. Arima, H. M. Rønnow, Y. Tokura, and Y. Taguchi, Skyrmion formation in a bulk chiral magnet at zero magnetic field and above room temperature, Phys. Rev. Matter. **1**, 074405 (2017).

[13] K. Karube, J. S. White, N. Reynolds, J. L. Gavilano, H. Oike, A. Kikkawa, F. Kagawa, Y. Tokunaga, H. M. Rønnow, Y. Tokura *et al.*, Robust metastable skyrmions and their triangular–square lattice structural transition in a high-temperature chiral magnet, Nat. Mater. **15**, 1237 (2016).

[14] O. Boulle, J. Vogel, H. Yang, S. Pizzini, D. de Souza Chaves, A. Locatelli, T. O. Menteş, A. Sala, L. D. Buda-Prejbeanu, O. Klein *et al.*, Room-temperature chiral magnetic skyrmions in ultrathin magnetic nanostructures, Nat. Nanotechnol. **11**, 449 (2016).

[15] S. A. Montoya, S. Couture, J. J. Chess, J. C. T. Lee, N. Kent, M. Y. Im, S. D. Kevan, P. Fischer, B. J. McMorran, S. Roy *et al.*, Resonant properties of dipole skyrmions in amorphous Fe/Gd multilayers, Phys. Rev. B **95**, 224405 (2017).

[16] S. A. Montoya, S. Couture, J. J. Chess, J. C. T. Lee, N. Kent, D. Henze, S. K. Sinha, M. Y. Im, S. D. Kevan, P. Fischer et al., Tailoring magnetic energies to form dipole skyrmions and skyrmion lattices, Phys. Rev. B **95**, 024415 (2017).

[17] I. Lemesh and G. S. D. Beach, Twisted domain walls and skyrmions in perpendicularly magnetized multilayers, Phys. Rev. B **98**, 104402 (2018).

[18] D. A. Gilbert, J.-W. Liao, B. J. Kirby, M. Winklhofer, C.-H. Lai, and K. Liu, Magnetic Yoking and Tunable Interactions in FePt-Based Hard/Soft Bilayers, Sci. Rep. **6**, 32842 (2016).

[19] N. Tang, J.-W. Liao, S.-T. Chui, T. Ziman, A. J. Grutter, K. Liu, C.-H. Lai, B. J. Kirby, and D. A. Gilbert, Controlling magnetic configuration in soft–hard bilayers probed by polarized neutron reflectometry, APL Mater. **10**, 011107 (2022).

[20] W. Legrand, J.-Y. Chauleau, D. Maccariello, N. Reyren, S. Collin, K. Bouzehouane, N. Jaouen, V. Cros, and A. Fert, Hybrid chiral domain walls and skyrmions in magnetic multilayers, Sci. Adv. **4**, eaat0415 (2018).

[21] H. S. Park, X. Yu, S. Aizawa, T. Tanigaki, T. Akashi, Y. Takahashi, T. Matsuda, N. Kanazawa, Y. Onose, D. Shindo et al., Observation of the magnetic flux and three-dimensional structure of skyrmion lattices by electron holography, Nat. Nanotechnol. **9**, 337 (2014).

[22] C. F. Majkrzak, Polarized neutron reflectometry, Physica B Condens. **173**, 75 (1991).

[23] S. Nouhi, M. S. Hellsing, V. Kapaklis, and A. R. Rennie, Grazing-incidence small-angle neutron scattering from structures below an interface, J. Appl. Crystallogr. **50**, 1066 (2017).

[24] M. C. Langner, S. Roy, S. K. Mishra, J. C. T. Lee, X. W. Shi, M. A. Hossain, Y. D. Chuang, S. Seki, Y. Tokura, S. D. Kevan et al., Coupled Skyrmion Sublattices in $Cu_2OSeO_3$, Physical Review Letters **112**, 167202 (2014).

[25] D. A. Gilbert, B. B. Maranville, A. L. Balk, B. J. Kirby, P. Fischer, D. T. Pierce, J. Unguris, J. A. Borchers, and K. Liu, Realization of ground-state artificial skyrmion lattices at room temperature, Nat. Commun. **6**, 8462 (2015).

[26] T. Kyrey, M. Ganeva, K. Gawlitza, J. Witte, R. von Klitzing, O. Soltwedel, Z. Di, S. Wellert, and O. Holderer, Grazing incidence SANS and reflectometry combined with simulation of adsorbed microgel particles, Physica B Condens. **551**, 172 (2018).

[27] H. J. C. Lauter, V. Lauter, and B. P. Toperverg, *Reflectivity, Off-Specular Scattering, and GI-SAS* (2012), Polymer Science: A Comprehensive Reference.

[28] J. F. Ankner and G. P. Felcher, Polarized-neutron reflectometry, J. Magn. Magn. Mater. **200**, 741 (1999).

[29] See Supplemental Material [link to be inserted by publisher] for the measured magnetic hysteresis loop.

[30] G. L. Squires, *Introduction to the Theory of Thermal Neutron Scattering* (Cambridge University Press, Cambridge, (2012), 3 edn.

[31] R. M. Moon, T. Riste, and W. C. Koehler, Polarization Analysis of Thermal-Neutron Scattering, Phys. Rev. **181**, 920 (1969).

[32] B. Maranville, W. Ratcliff II, and P. Kienzle, reductus: a stateless Python data reduction service with a browser front end, J. Appl. Crystallogr. **51**, 1500 (2018).

[33] OOMMF User's Guide, Version 1.0

[34] D. A. Gilbert, A. J. Grutter, P. M. Neves, G.-J. Shu, G. Zimanyi, B. B. Maranville, F.-C. Chou, K. Krycka, N. P. Butch, S. Huang et al., Precipitating ordered skyrmion lattices from helical spaghetti and granular powders, Phys. Rev. Matter. **3**, 014408 (2019).

[35] R. C. Taylor and A. Gangulee, Magnetic anisotropy in evaporated amorphous films of the ternary system $Gd_x(Fe_{1-y}Co_y)_{1-x}$, J. Appl. Phys. **48**, 358 (1977).

[36] C. F. Majkrzak, C. Metting, B. B. Maranville, J. A. Dura, S. Satija, T. Udovic, and N. F. Berk, Determination of the effective transverse coherence of the neutron wave packet as employed in reflectivity investigations of condensed-matter structures. I. Measurements, Phys. Rev. A **89**, 033851 (2014).

[37] A. Singh, M. K. Sanyal, J. C. T. Lee, J. J. Chess, R. Streubel, S. A. Montoya, M. K. Mukhopadhyay, B. J. McMorran, E. E. Fullerton, P. Fischer et al., Discretized evolution of solitons in the achiral stripe phase of a Fe/Gd thin film, Phys. Rev. B **105**, 094423 (2022).

[38] D. A. Gilbert, L.-W. Wang, T. J. Klemmer, J.-U. Thiele, C.-H. Lai, and K. Liu, Tuning magnetic anisotropy in (001) oriented L10 $(Fe_{1-x}Cu_x)_{55}Pt_{45}$ films, Appl. Phys. Lett. **102**, 132406 (2013).

[39] M. D. Kuz'min, Shape of Temperature Dependence of Spontaneous Magnetization of Ferromagnets: Quantitative Analysis, Phys. Rev. Lett. **94**, 107204 (2005).

# Figures
## Fig. 1

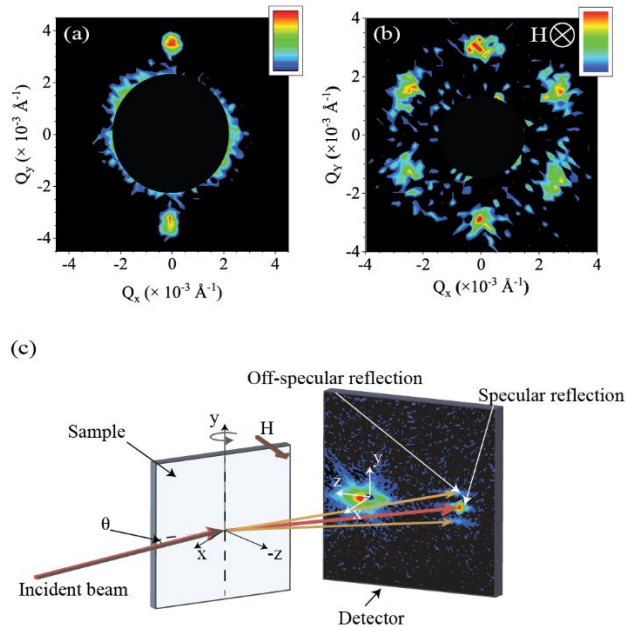

**Figure 1** Measured SANS signal in (a) the stripe state and (b) the skyrmion state, both taken in a small, residual field, aligned with the neutron beam along the film normal direction. The $z$-contrast uses a logarithmic scale. (c) Diagram of the GISANS configuration with the red arrow indicating the specular reflection and orange arrows indicating the off specular (GISANS) peaks. Coordinates for the GISANS results are defined relative to the center of the direct beam. For GISANS measurements, the sample is rotated around its y-axis.

**Fig 2**

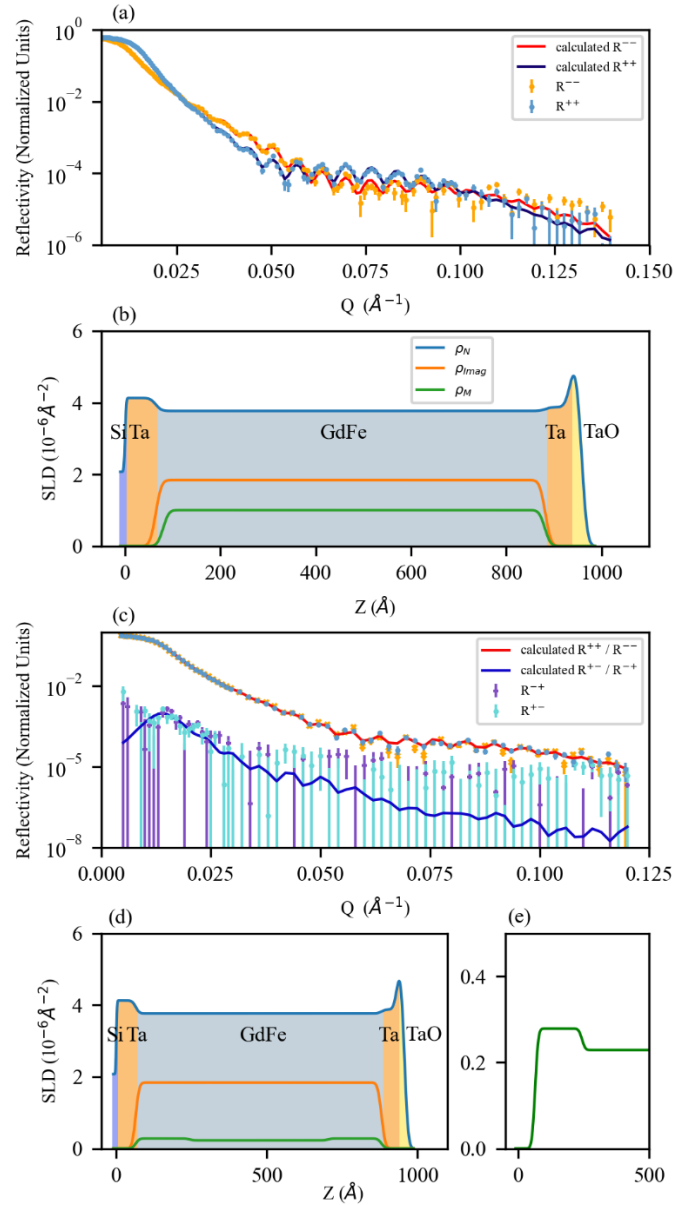

**Figure 2** (a) Specular reflectivity measurement and (b) converged depth profile for the Gd/Fe multilayer in the saturated state (in-plane saturation, 500 mT). The neutron polarization axis is in-plane and parallel to the saturating field in this scattering configuration. Only the non-spin flip data is shown since there is no appreciable spin-flip signal in saturation. (c) Specular reflectometry and (d) converged depth profile measured in a 1 mT field, in the skyrmion state with the magnetic surface state emphasized in (e). For (a) the neutron polarization axis is in the plane of the film while for (c) the neutron polarization axis is out of plane and parallel to the small guide field.

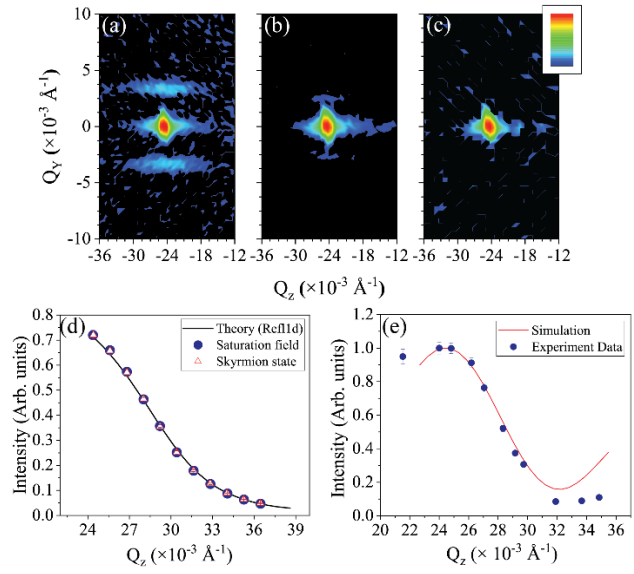

**Figure 3** GISANS data in (a) the stripe state, (b) skyrmion state, and (c) saturated state. The *z*-contrast uses a logarithmic scale. For all three measurements the specular peak appears at $Q_Y = 0$, while the off-specular peaks appear at $Q_y = \pm 0.00338$ Å$^{-1}$ and $Q_y = \pm 0.0025$ Å$^{-1}$ for the stripe and skyrmion states, respectively. (d) Integrated intensity of the specular reflection from GISANS for the skyrmion (triangle), out-of-plane saturated states (circle), and theory from the Refl1d model (line). (e) Integrated intensity from the off-specular diffraction peak from GISANS (circle) and theoretical results from micromagnetic modeling (line).

**Fig 4**

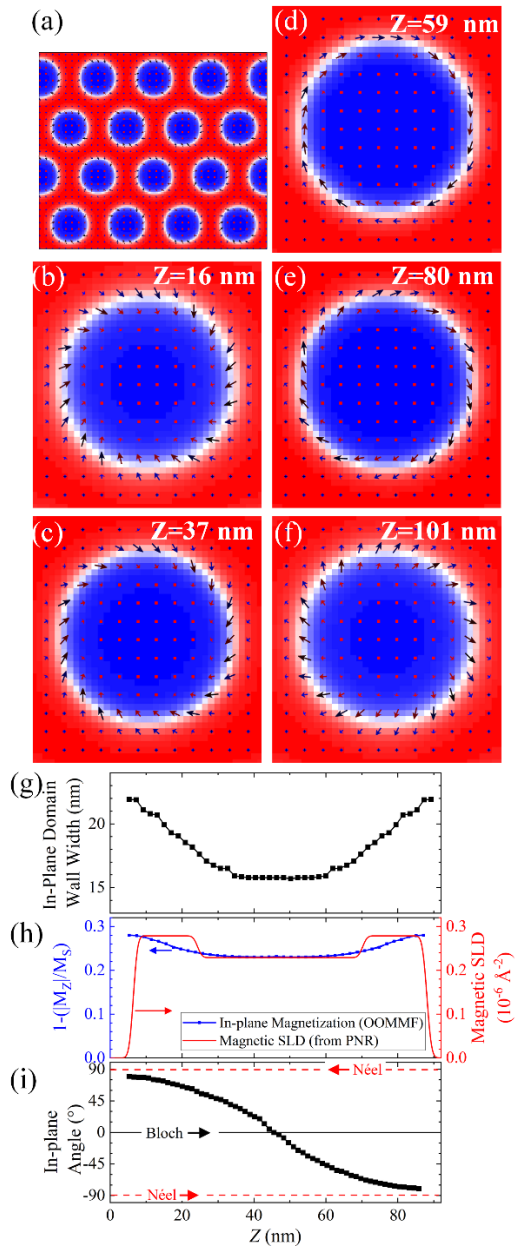

**Figure 4** (a) Plane view image of the simulated skyrmion lattice. (b)-(f) Slices of the magnetic configuration taken along the length of the skyrmion tube, with (b) Z=16 nm corresponding to the top surface and an inward wrapping Néel cap, (d) Z=59 nm corresponding to the middle and a clockwise Bloch type, and (f) Z=101 nm corresponding to the bottom surface and an outward wrapping Néel cap. (g) The in-plane width of the skyrmion boundary region. (h) The total in-plane magnetization from OOMMF integrated versus depth and compared to the magnetic SLD profile from the PNR results. (i) The in-plane angle of the skyrmion boundary region versus depth, with +90°/0/-90° corresponding to inward Néel, clockwise Bloch, and outward Néel, respectively. Note that 0° is defined as the azimuthal direction.

**Fig 5**

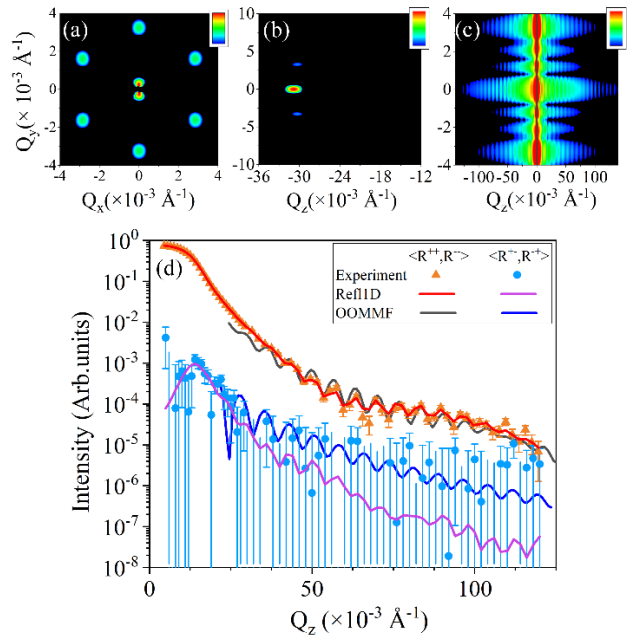

**Figure 5** Calculated 2D (a) transmission SANS, (b) GISANS measurements simulated at $Q_z = -0.031$ Å$^{-1}$, and (c) cumulative GISANS pattern from the OOMMF simulations of the skyrmion state. The *z*-contrast uses a logarithmic scale. The cumulative GISANS pattern is the sum of all measured theta angles (i.e., all measured $Q_z$ values), while the experimental results are measured at sequential angles, analogous to panel (b) and Figure 3(b). (d) Average non-spin-flip specular reflectivity $<R^{++}, R^{--}>$ and spin-flip reflectivity $<R^{+-}, R^{-+}>$ from PNR, OOMMF (1D depth profile) and three-dimensional micromagnetic simulation (OOMMF, shown in Figure 4) at $Q_y = 0$. A constant background was added to the three-dimensional OOMMF model.